\documentclass[prab,graphicx,preprint]{revtex4-1}   	
\usepackage{geometry}                		
\usepackage{graphicx}				
\usepackage{setspace}								
\usepackage{amssymb}
\usepackage{amsmath}   
\usepackage{subcaption}

\begin{document}

\title{Current transmission and nonlinear effects in un-gated thermionic cathode RF guns}

\author{J. P. Edelen}
\thanks{jedelen@fnal.gov}
\affiliation{Fermi National Accelerator Laboratory, Batavia IL, 60510, USA}
\thanks{Operated by Fermi Research Alliance, LLC under Contract No. De-AC02-07CH11359 with the United States Department of Energy}
\author{J. R. Harris} 
\affiliation{Directed Energy Directorate, Air Force Research Laboratory, Albuquerque NM, 87117, USA}

\begin{abstract}
Un-gated thermionic cathode RF guns are well known as a robust source of electrons for many accelerator applications. These sources are in principle scalable to high currents without degradation of the transverse emittance due to control grids but they are also known for being limited by back-bombardment. While back-bombardment presents a significant limitation, there is still a lack of general understanding on how emission over the whole RF period will affect the nature of the beams produced from these guns. In order to improve our understanding of how these guns can be used in general we develop analytical models that predict the transmission efficiency as a function of the design parameters, study how bunch compression and emission enhancement caused by Schottky barrier lowering affect the output current profile in the gun, and study the onset of space-charge limited effects and the resultant virtual cathode formation leading to a modulation in the output current distribution. 

\end{abstract}
\maketitle

\section{Introduction}
Electron beams for use in the next generation of light sources are predominately created by photocathode radio frequency (RF) guns [1,2,3,4]. While photocathode RF guns are considered to be a reliable source of high-quality beams, they require high-average-power drive lasers which tend to be large, expensive, and operationally labor intensive. Additionally, these guns require the use of high-quantum-efficiency cathodes that typically have a limited lifetime (e.g. hours to days) and require frequent replacement and/or rejuvenation [5, 6]. These issues could be avoided by using thermionic cathodes which are robust and long-lived [7]. Because the current emitted from a thermionic cathode depends on its temperature (which cannot be rapidly changed due to the cathode's thermal mass), emission is typically gated using a control grid [8, 9, 10]. While they are effective for this purpose, these grids tend to degrade the transverse beam emittance, especially at high current. Techniques for laser gating, where a laser pulse is used to heat the surface of the cathode and produce thermionic emission, have also been investigated [11, 12]. For high currents this technique becomes impractical, as the cooling time of the cathode surface will be longer than that of a single RF period. A scheme where the emission is gated by the RF gun cavity field alone is attractive because in principle it allows for unobstructed increase in beam current without degradation of the transverse emittance. Additionally, it would not require the use of high-power drive lasers and supplies consistent, reliable emission from a long-lasting source. 

One complication of this approach is that electrons are emitted continuously over the RF period. As a result the beam often requires further compression, for instance by an alpha magnet, before being injected into further acceleration stages [13]. Additionally, some electrons emitted late relative to the RF will not gain enough energy to exit the cavity and will be accelerated back to the cathode [14]. These back-bombarded electrons can have adverse effects on the machine performance [15] and even damage the cathode [16]. The understanding of back-bombardment as a limitation for these guns has been extensively studied, and many back-bombardment mitigation techniques exist. However, there is a dearth of understanding about how emission over the whole RF period will impact the beam properties for this class of electron sources.  Because the electrons are accelerated in an RF field from near rest, particles emitted early in the RF period will see a large time-varying field and will subsequently undergo significant bunch compression compared with particles emitted later in the RF period [17, 18]. This will lead to a nonlinear current profile  along the beam [19]. A further complication of the RF field on the cathode is that increased emission due to Schottky barrier lowering will impart an additional nonlinearity onto the emitted current that cannot be compensated for downstream. These uncompensated current nonlinearities can result in stronger CSR forces in compression chicanes [20] leading to further reduced performance. Additionally, the onset of space-charge-limited emission will introduce a virtual-cathode-like depression which can modulate the field on the cathode, resulting in a highly structured beam being emitted from the gun. 

In this paper we seek to characterize, in a general sense, the nonlinear current profiles produced by these types of guns and examine how the geometry, field, and injected current impact the output current of the beam. This begins with an analysis of the transmission efficiency of the gun, followed by a discussion of the effects of bunch compression and Schottky barrier lowering on the output current profile. Finally, we discuss the effects of increasing the current to the point of being space-charge-limited and how this will impact the output beam profile. For simplicity, the class of guns considered in this paper are all single-cell, with gap length being treated as an adjustable parameter.

\section{Transmission efficiency}  
Before attempting to understand the details of the beam profile we first investigate how much of the emitted current we can expect to extract, the transmission efficiency, for various design parameters. In order to estimate the transmission efficiency we calculate the boundary between particles that are back-bombarded and particles that are transmitted through the gun. This is accomplished using the effective transit time of particles in the gap defined by $t_\mathrm{transit} = {\lambda \over \alpha v_\mathrm{eff}}$ [14], where $v_\mathrm{eff}$ is the effective velocity given by $v_\mathrm{eff}= {c}\sqrt{1-\left(1+{qE_\mathrm{peak}\lambda\over 2m_0 c^2\alpha}\right)^{-2}}$, $\lambda$ is the RF wavelength, $\alpha$ is the fractional gap length where $L_\mathrm{gap} = \lambda/\alpha$, $E_\mathrm{peak}$ is the peak field in the cavity, $m_0$ is the electron mass, and $c$ is the speed of light.  The boundary between particles that are emitted and particles that are back-bombarded is defined as $t_\mathrm{bb} = \tau/2-t_\mathrm{transit}/2$ [ibid], where $\tau$ is the RF period. Taking the ratio of the back-bombardment boundary to the total emission time, $2 t_\mathrm{bb} / \tau$, yields the transmission efficiency given by Equation 1. 
\begin{equation} 
e = 1 - {c\over \alpha v_\mathrm{eff}}
\end{equation} 
Note that Equation 1 assumes that all particles either exit the gun or strike the cathode within one RF period. In order to assess the validity of Equation 1, we numerically computed the extraction efficiency using a 1-D particle pusher for guns with a frequency ranging from 650 MHz to 3.9 GHz, a peak field ranging from 1 MV/m to 160MV/m and a value of $\alpha$ ranging from 3 to 14. Here $\alpha = 3.33$ is typical for photocathode-guns with a long cathode cell to perform emittance compensation and $\alpha = 14$ is approaching the upper limit of where guns can be practically manufactured at these frequencies. Allowing the peak field to increase above the break-down limits gives asymptotic estimates for the transmission efficiency. Across this range of gun designs, Equation 1 predicted the extraction efficiency to within 9\% RMS. Figure 1 shows a representative comparison of the analytic model to the simulations. 

\begin{figure}[h]
\centering
\includegraphics[width=0.5\textwidth]{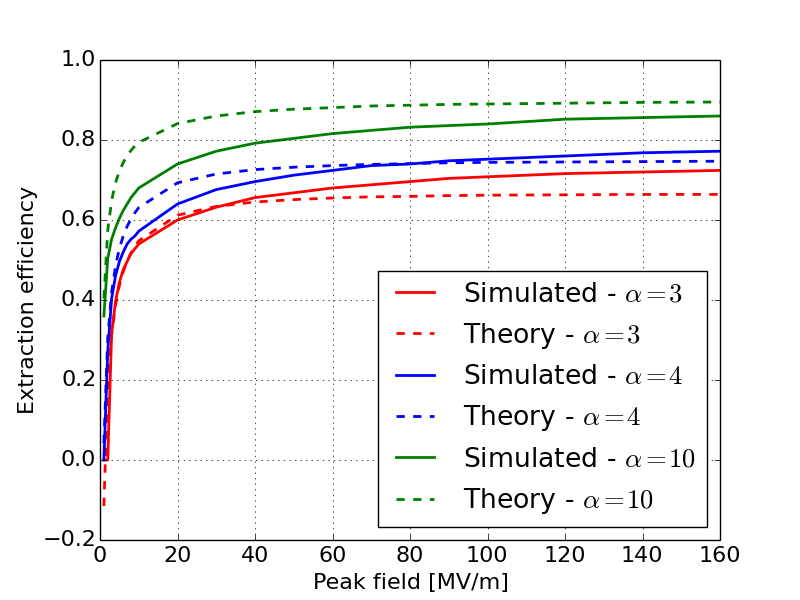}
\caption{Comparison of theory and simulation for the extraction efficiency at three different values of $\alpha$ at an RF frequency of 1.3GHz. }
\end{figure}

Here we see that the analytic model is slightly over predicting the influence of the gap length on the transmission efficiency. In spite of this, the relative shape of Equation 1 agrees well with the simulations. 

\section{Current distribution resulting from bunch compression} 
While the transmission efficiency is a useful metric for gun design we are most concerned with the fine structure of the output current profile produced by this class of guns. Because of the nonuniform acceleration (time varying field), electrons emitted early during the RF period will take longer to accelerate than electrons emitted closer to crest. This will result in bunch compression that varies throughout the RF period and will ultimately introduce nonlinearities in the output current distribution. While this process has been described in detail for photocathode RF guns [17, 18], it has not been fully characterized for a broad range of thermionic RF guns. In order to better understand how bunch compression will affect the output current, we examine the output phase of electrons produced in the gun as a function of the input phase [18]:

\begin{equation} 
\phi(\tilde\gamma) = {1 \over 2\alpha_\mathrm{rf} \sin\left(\phi_\mathrm{eff}\right)} \left( \sqrt{\tilde\gamma^2 - 1} - \left( \tilde\gamma - 1 \right)\right) + \phi_0.
\end{equation} 

Here $\alpha_\mathrm{rf} = qE_{\mathrm{peak}}/(2m_0c^2k)$, $\phi_\mathrm{eff} = \phi_0 + 1/(2\alpha_\mathrm{rf})$, and $\tilde\gamma = 1 + 2\alpha_\mathrm{rf} k \sin\left(\phi_\mathrm{eff}\right)z$, where $k = 2\pi/\lambda$, $\phi_0$ is the injection phase, and $z$ is the longitudinal position along the gun. For a single cell gun with a fixed geometry we can make the substitution $z = \lambda / \alpha$ and then reduce Equation 3 to an expression that is only a function of the gun parameters and the input phase, Equation 3. 

\begin{equation} 
\phi(\phi_0)= {2\pi \over \alpha } \left(\sqrt{1 +  {\alpha \over 2\pi \alpha_\mathrm{rf} \sin\left(\phi_\mathrm{eff}\right)}} - 1\right)  + \phi_0 
\end{equation} 

Taking the derivative of Equation 3 with respect to the input phase results in an expression that describes the level of compression present for a given input phase, Equation 4. This is similar to the bunch compression factor derived for photocathode RF guns [17, 18] with the key difference that it is generalized for different cathode-cell gap-lengths. Additionally Equation 4 only describes the beam at the exit of the cathode cell (i.e. it is not asymptotic)   

\begin{equation} 
{d\phi \over d\phi_0} = 1-{ \left(\cos(\phi_0 + 1/(2\alpha_\mathrm{rf})) \over \sin^2(\phi_0 + 1/(2\alpha_\mathrm{rf}))\right) \over 2 \alpha_\mathrm{rf} \sqrt{ {\alpha \over 2\pi\alpha_\mathrm{rf} \sin(\phi_0 + 1/(2\alpha_\mathrm{rf}))} + 1}}.
\end{equation} 

The current distribution along the bunch will scale inversely to the bunch compression factor (Equation 4). Figures 2 and 3 show the predicted current distribution as a function of input phase for different values of peak field and $\alpha$ respectively. Note that Equation 4 will produce non-physical results when the bunch compression factor is less then zero or when the denominator of Equation 4 results in a complex number. In order to remove these for visualization purposes and for comparison, the phase on each curve is shifted such that the maximum current appears at an input phase of zero. Additionally, the vertical scale was restricted to see the peak at the head of the bunch and the tails. As we will see later, these results are consistent with simulation results. Figure 2 shows that as the field in the gap increases, the peak in the current becomes more prominent and the current at the center of the bunch becomes more uniform. 

\begin{figure}[h]
\centering
\includegraphics[width = 0.5\textwidth] {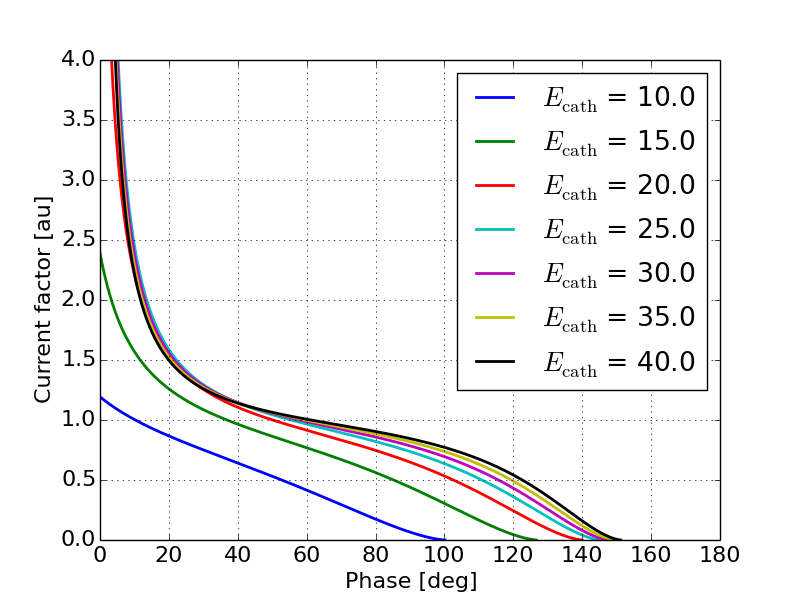}
\caption{Current factor as a function of input phase for several values of peak field. For these curves $\alpha$ was set to 3.33} 
\end{figure} 

\begin{figure}[h]
\centering
\includegraphics[width = 0.5\textwidth] {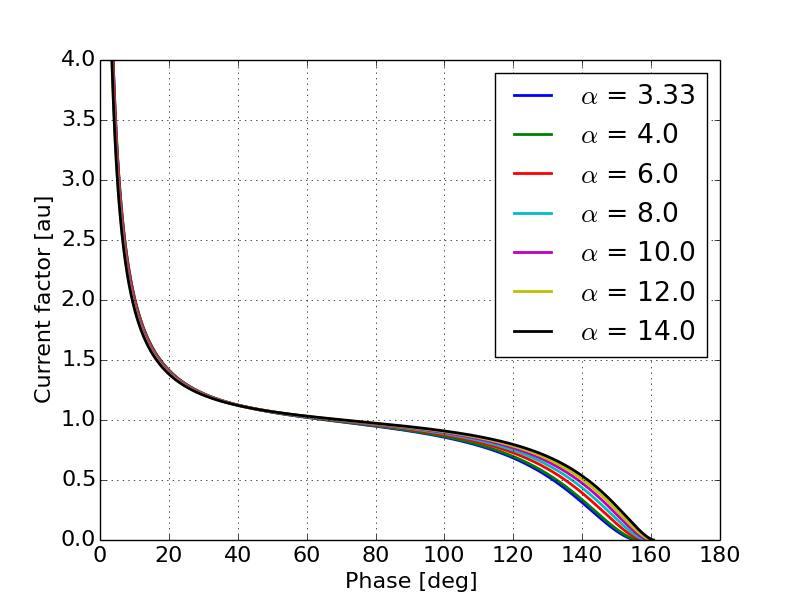}
\caption{Effective current as a function of input phase for several values of $\alpha$. For these curves $E_0$ was set to 30 MV/m} 
\end{figure} 

Figure 3 shows that as $\alpha$ increases the bunch current is in general more uniform along the middle of the bunch with smaller tails. This can be explained by a decreased peak energy due to the shorter gap. Electrons emitted on crest will see a quicker acceleration but will leave the gun quicker and therefore reduce the total amount of compression. In order to verify the results in Figures 2 and 3 1-D simulations were performed using a particle pusher. Figures 4 and 5 show the current as a function of phase for different values of peak field and $\alpha$ respectively. Note that the analytical expressions are derived in terms of the input phase which is appropriate for photo-cathode guns due to the need to synchronize the laser pulse with the RF. While this does provide intuition for what to expect from the output beam in a thermionic gun, the theory is not well suited for predicting the output beam profile with respect to the output phase. Because we are concerned with the details of the output beam, simulation results shown in this paper are presented with respect to the output phase rather than the input phase.

\begin{figure}[h]
\centering
\includegraphics[width = 0.5\textwidth] {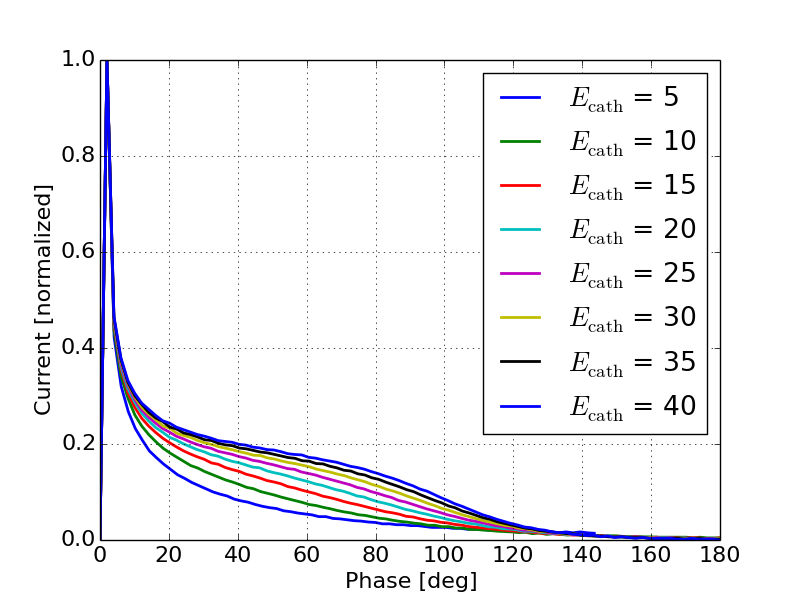}
\caption{Normalized current as a function of output phase for several values of peak field from 1-D simulations. For these curves $\alpha$ was set to 3.33} 
\end{figure}

\begin{figure}[h]
\centering
\includegraphics[width = 0.5\textwidth] {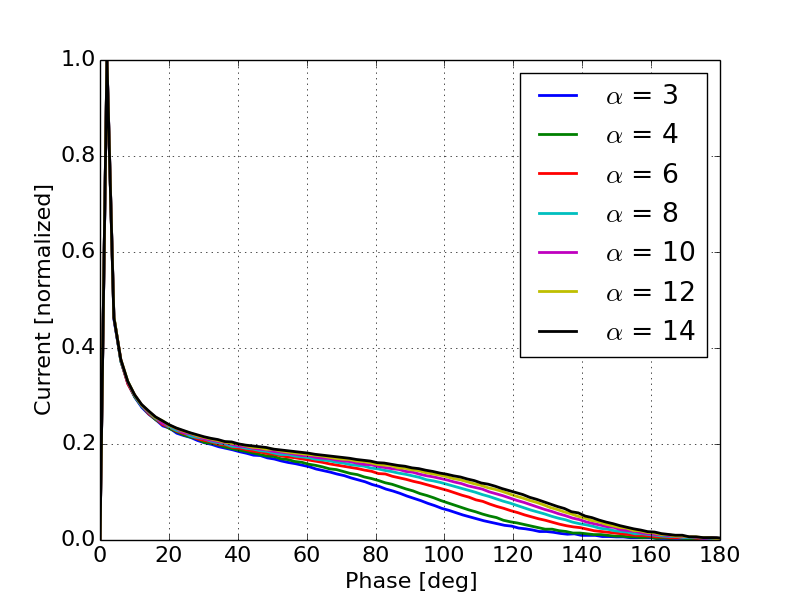}
\caption{Normalized current as a function of output phase for several values of $\alpha$ from 1-D simulations. For these curves $E_0$ was set to 30 MV/m} 
\end{figure} 

Comparing Figures 2 and 3 to Figures 4 and 5 respectively shows that in general the theory captures the relative behavior of the output current as a function of the gun parameters. In general the theory tends to exaggerate the impact of both peak field and $\alpha$.  While bunch compression introduces a significant nonlinearity in the beam current distribution, the Schottky effect plays an equally significant role in the current profile and can greatly increase the nonlinearities in the current distribution.

\section{Effects of Schottky barrier lowering on current distribution} 
Schottky barrier lowering will have two primary effects on the output current: first is it will increase the average current, and second is it will modulate the cathode emission during the RF period embedding the RF waveform onto the emitted beam current profile. 

In order to understand exactly how this manifests itself we examine the Richardson-Dushman equation in the presence of transient RF fields (Equation 5). In this section we assume that the space-charge field is negligible and therefore the field on the cathode is determined completely by the applied RF field. 

\begin{equation} 
J_\mathrm{emit} = A T^2 \exp\left[- {W - \sqrt{q^3E_\mathrm{peak}\sin\left(\phi_0\right)\over 4\pi\epsilon_0} \over k_b T}\right]
\end{equation} 

Here $J_\mathrm{emit}$ is the emitted current density, $A$ is an emission constant, $T$ is the cathode temperature, $W$ is the cathode work function, $q$ is the electron charge, $\epsilon_0$ is the permittivity of free space, and $k_b$ is the Boltzman constant. Figure 6 shows the emission curve of a thermionic cathode as a function of phase for several values of peak field. Note that for this figure the design (i.e. zero-field) output current was 100 mA. Even when the field on the cathode is as low as 10 MV/m the peak emission was three times greater than the design current. 

\begin{figure}[h]
\centering
\includegraphics[width = 0.5\textwidth] {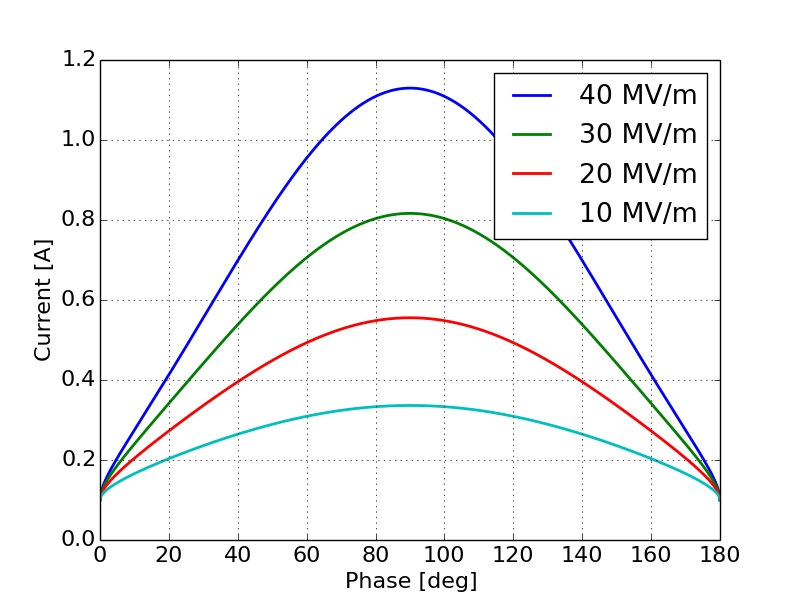}
\caption{Emission curve as a function of phase for several values of peak field with a design current of 100 mA.} 
\end{figure} 

One can estimate how he Schottky effect will alter the average emitted current by integrating Equation 5 with respect to $\phi$. However, because the integral cannot be solved analytically the average current can be approximated by substituting the average field in the gap, $E_\mathrm{ave}$, for the time varying field in the gap, $E_\mathrm{peak}\sin(\phi)$. Here $E_\mathrm{ave} = 2E_\mathrm{peak}/\pi$. For a temperature range from 1000 K to 2000 K and a field range from 1 MV/m to 100 MV/m, the root mean squared error between numerically integrating Equation 5 and using the average field for computing the average emitted current, was 5\%. If one uses the peak field in the gap to compute the average emitted current, the calculated average emitted current will be off by a factor of two from the true average emitted current, on average. In order to ensure an accurate estimate of the average emitted current from this type of gun one should use the average field on the cathode to account for the Schottky effect. This can be combined with the transmission efficiency defined in Section II to estimate the average output current of the gun for a given design. 

In order to understand how the time varying emission enhancement will impact the current distribution, we combine Equations 4 and 5 by treating the bunch compression factor as a transfer function on the emission, Equation 6. 
\begin{equation} 
J_\mathrm{out} = J_\mathrm{emit} {1\over d\phi/d\phi_0}
\end{equation} 
This will take into account both the bunch compression and the time varying emission enhancement due to the Schottky effect. Figures 7 and 8 show example current distributions using Equation 6 for a constant peak field and constant $\alpha$ respectively. As with Figures 2 and 3, the current distribution is given as a function of the input phase. 

\begin{figure}[h]
\centering
\includegraphics[width=0.5\textwidth]{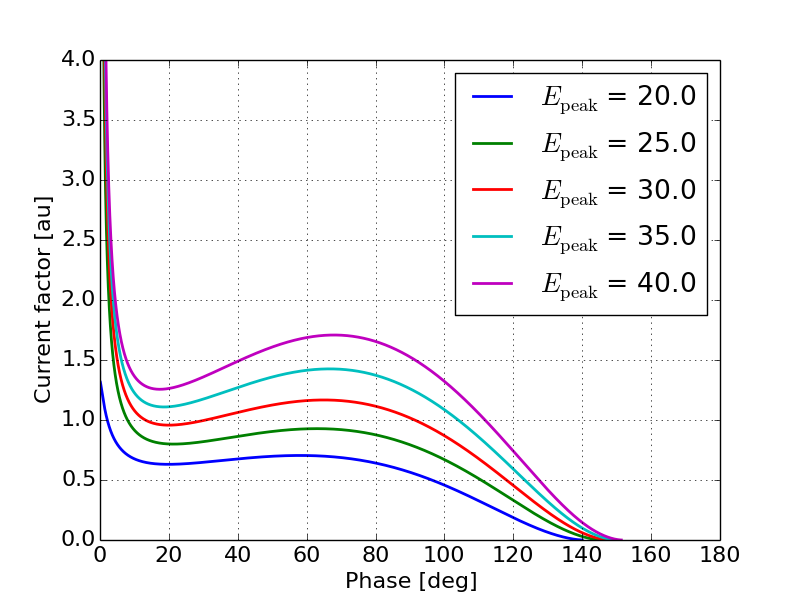}
\caption{Effective current due to emission enhancement as a function of input phase for a several peak fields. For these curves $\alpha$ is 3.33. } 
\end{figure}

\begin{figure}[h]
\centering
\includegraphics[width=0.5\textwidth]{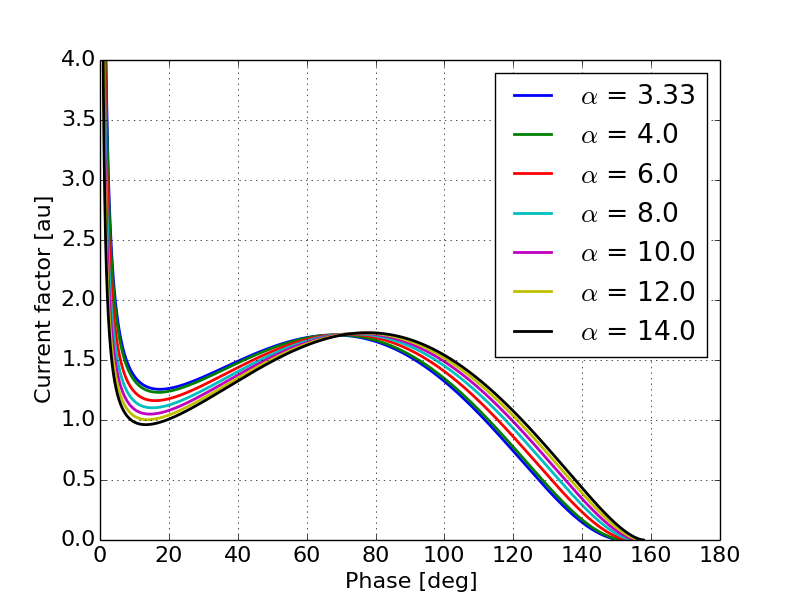}
\caption{Effective current due to emission enhancement as a function of input phase for a several values of $\alpha$. For these curves $E_\mathrm{cath} $ is 30 MV/m.} 
\end{figure}

Figures 7 and 8 give some understanding of the relationship between the current distribution and the peak field as well as the gap length. Figure 7 shows the somewhat trivial results that as the peak field increases the magnitude of the emission enhancement also increases. Figure 8 shows the relationship between the output current distribution and the gap length. It is expected that as $\alpha$ increases so will the amount of the RF curvature seen in the output beam as there is less overall compression in the beam current. 

Simulations using a 1-D particle pusher including the change to the emitted current as a function of the RF phase provide additional insight as to how this effect will change the output current profile. Figures 9 and 10 show the output current as a function of output phase for various values of peak field and various values of $\alpha$ respectively. 

\begin{figure}[h]
\centering
\includegraphics[width=0.5\textwidth]{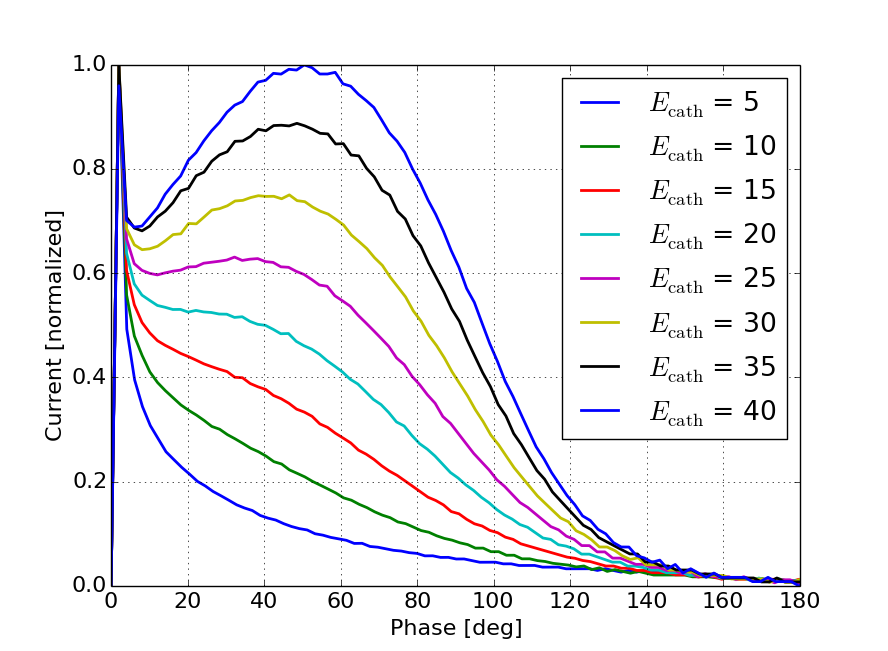}
\caption{Effective current due to emission enhancement as a function of output phase for a several peak fields from 1-D simulations. For these curves $\alpha$ is 3.33.}
\end{figure}

\begin{figure}[h]
\centering
\includegraphics[width=0.5\textwidth]{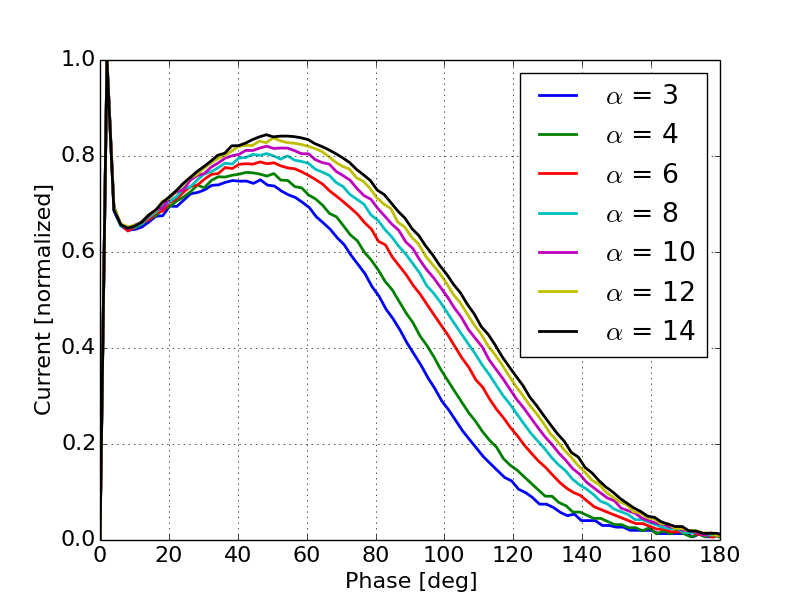}
\caption{Effective current due to emission enhancement as a function of output phase for a several values of $\alpha$ from 1-D simulations. For these curves the peak field was 30 MV/m} 
\end{figure} 

Figures 9 and 10 support the basic trends predicted by the analytical calculations. Figure 9 shows that as the peak field increases, the output current tends to more closely follow the RF waveform. This can be explained by the fact that the variation in the emitted current dominates over the bunch compression. From Figure 6 we can see that for a gun with a peak field of 40 MV/m the emitted current will vary by a factor of two between 30 and 90 degrees while the bunch compression in this regime only changes by about 10\%. Additionally, as the peak field decreases the shape of the output current tends to follow the curves shown in Figure 4 more closely indicating that in the lower field regime bunch compression plays a larger role in determining the output current. Figure 10 shows that as the gap length decreases we see more of the RF waveform in the output current. This is again due to the fact that there is less energy gain from the shorter gap and therefore bunch compression is less significant, allowing the emission enhancement effects to dominate the output current distribution.

\section{Space-charge effects} 
Sections III and IV have shown that due to emission occurring continuously over half the RF period there will be significant nonlinearities introduced in the output current distribution caused by both bunch compression and Schottky emission. When considering the class of guns for high current application it is important to also understand how the presence of significant space-charge effects will impact the current distribution. The onset of space-charge effects is first seen by Coulomb repulsion decreasing the net effect of bunch compression. This is followed by the space-charge field partially cancelling the field on the cathode thereby reducing the effect of emission enhancement on the output current. As the intensity of space-charge increases the gun eventually becomes fully space-charge limited resulting in the formation of a virtual cathode. Virtual cathode oscillations will modulate the field on the cathode leading to modulation of the emission. The combination of the virtual cathode and velocity bunching caused by low fields on the cathode results in a current modulation on the output beam.  In this section we will characterize the effect of space-charge beginning with the initial onset of space-charge, followed by a transition regime where the space-charge forces are significant and greatly impact the current distribution, to the fully space-charge-limited regime. 

 \subsection{Overview of simulations} 
 The simulations for this section were all performed using SPIFFE [21]. SPIFFE is a 2.5-D particle-in-cell electromagnetic field solver. The simulations were performed using an RF field with super-gaussian (Equation 7) longitudinal profile which serves as a good approximation of most RF cavities. 
 \begin{equation} 
 E_z(z) = \exp{\left(-\left(4\alpha z\over3\lambda\right)^4\right)}
 \end{equation}
 The simulations were all performed at an RF frequency of 1.3 GHz with an applied external magnetic field of 1 T. The external field serves to remove transverse motion from the simulation in order to concentrate on longitudinal space-charge effects.
 
\subsection{Onset of space-charge effects} 
We begin to see the effects of space-charge at relatively low currents. Figure 11 shows the current profile for an average injected current increasing from 100 mA to 1A. For comparison purposes each curve is normalized to its respective peak. Here we see that while the relative peaks in the current distribution do not change the relative minimum at the head of the bunch does. This indicates that the field generated by the space-charge is not enough to significantly reduce the field on the cathode and thereby reduce the impact of emission enhancement, but it is large enough to have an effect on the output current. This change in the minimum can be explained by a reduction in the effective bunch compression caused by Coulomb repulsion between the "slices" in the beam. At higher current the longitudinal profile of the beam will be less affected by strong bunch compression forces resulting in a slightly more uniform current profile.

\begin{figure}[h]
\centering
\includegraphics[width=0.5\textwidth]{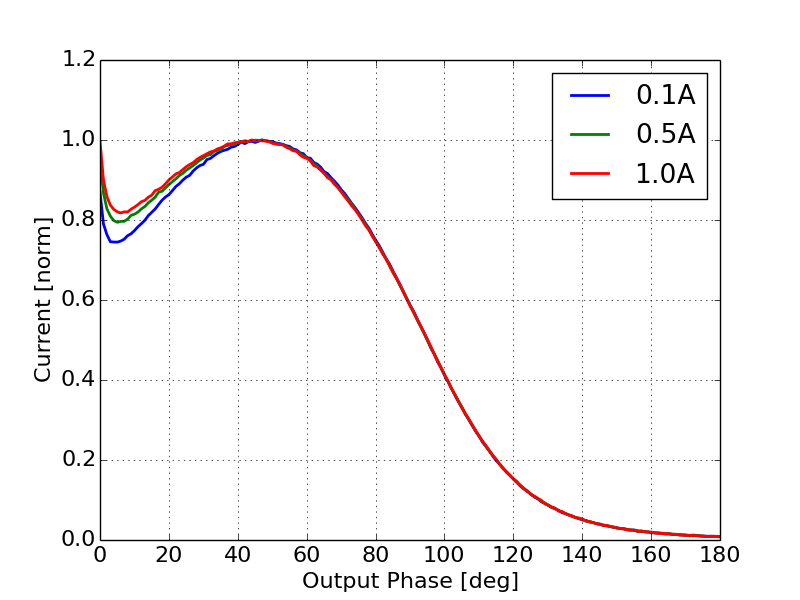}
\caption{Effective current as a function of output phase normalized to the peak current for several values of average injected current ranging from 100mA to 1A. For these curves $E_\mathrm{peak}$ was 40 MV/m and $\alpha$ is 3.33.}
\end{figure} 

As we increase the current, the fields generated by space-charge become significant enough to cancel out the field on the cathode and therefore reduce the impact of emission enhancement on the output current. Figure 12 shows the effective current as a function of output phase for injected currents ranging from 5 A to 50 A. Here we see that as the current increases the peak at the center of the current profile resulting from the Schottky effect becomes less prominent which is consistent with a reduction in emission enhancement shown in Figure 9. 

\begin{figure}[h]
\centering
\includegraphics[width=0.5\textwidth]{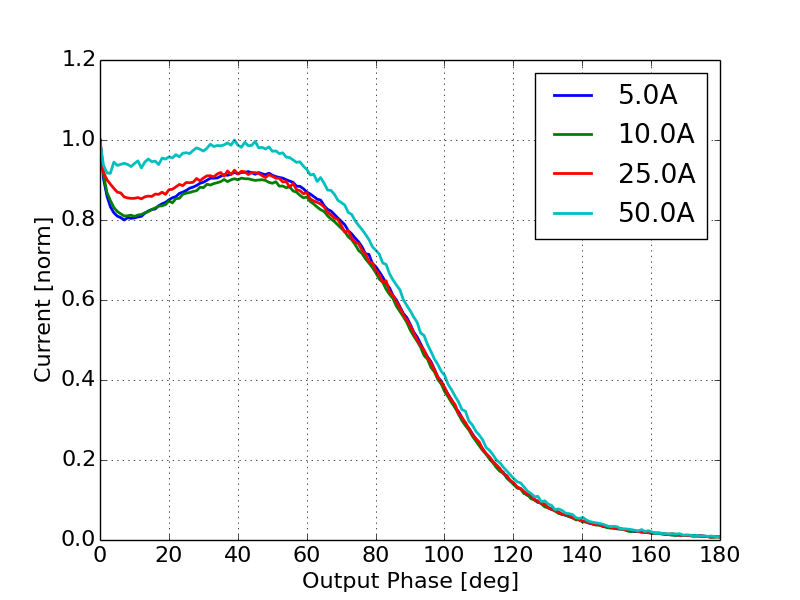}
\caption{Effective current as a function of output phase normalized to the peak current for average injected currents ranging from 5A to 50A. For these curves  $E_\mathrm{peak}$ was 40 MV/m and $\alpha$ is 3.33.}
\end{figure} 

These effects are confirmed by inspection of both the space-charge induced field on the cathode and the total field on the cathode as a function of phase during the simulations, Figures 13 and 14 respectively. SPIFFE allows for the saving of space-charge fields separately from other fields in the simulation, thus allowing us to study the space-charge field separate from the total field on the cathode. For currents less than 1 A the field on the cathode due to space-charge is too small to have a strong effect on the emission enhancement. As the current increases however the space-charge field begins to significantly cancel out the applied field on the cathode. Note that space-charge induced field on the cathode is not a pure sinusoid. Therefore we can also expect some change to the shape of the current caused by a change in shape of the emission enhancement away from a pure sinusoid. 

\begin{figure}[h]
\centering
\includegraphics[width=0.5\textwidth]{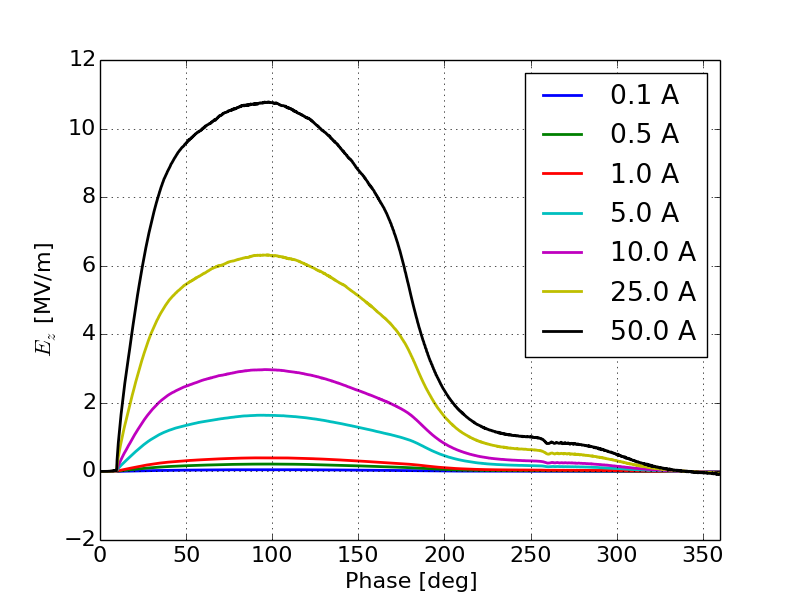}
\caption{Average field on the cathode due to space-charge as a function of phase for several values of average injected current.  $E_\mathrm{peak}$ was 40 MV/m and $\alpha$ is 3.33.}
\end{figure}

\begin{figure}[h]
\centering
\includegraphics[width=0.5\textwidth]{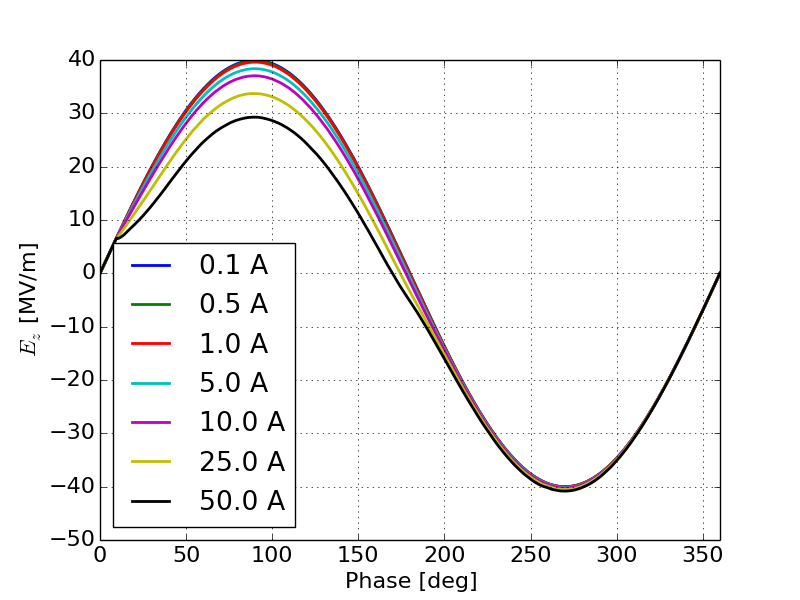}
\caption{Net effective field on the cathode due to space-charge as a function of phase for several values of average injected current.  $E_\mathrm{peak}$ was 40 MV/m and $\alpha$ is 3.33.}
\end{figure}

\subsection{Intense space-charge effects}
As the injected current increases we anticipate a continued suppression of the emission enhancement effect until we begin to see the onset of space-charge limited emission. In DC guns, space-charge limited emission is accompanied by the formation of a virtual cathode depression causing some electrons to be reflected back to the cathode. As this injected current increases, we begin to see a modulation of the virtual cathode [22]. In RF guns we expect to see a similar virtual cathode depression forming especially at very high currents [23]. In order to study how the onset of space-charge limited emission will impact the output current we increased the injected current from 50 A to 1000 A. Figure 15 shows the output beam current as a function of phase as the gun transitions from the regime where space-charge canceled some of the emission enhancement effect to the regime where emission is fully space-charge limited. 

Figure 15a shows the initial onset of space-charge limited effects that brings with it a modulation in the beam current profile. This structure is caused by virtual cathode oscillations described in detail later. Figure 15b shows the gun becoming fully space-charge limited as the injected current increases but the envelope of the transmitted current no longer increases proportionally to the injected current. Additionally, as the injected current increases the modulation in the output current profile continues to become more pronounced. Figure 15c shows the gun becoming fully space-charge limited with no change in the envelope of the current profile, but there continue to be subtle changes in the intensity and frequency of the modulation on the output current.

 \begin{figure}[h]
    \centering
    \begin{subfigure}[]{0.5\textwidth}
        \includegraphics[width=\textwidth]{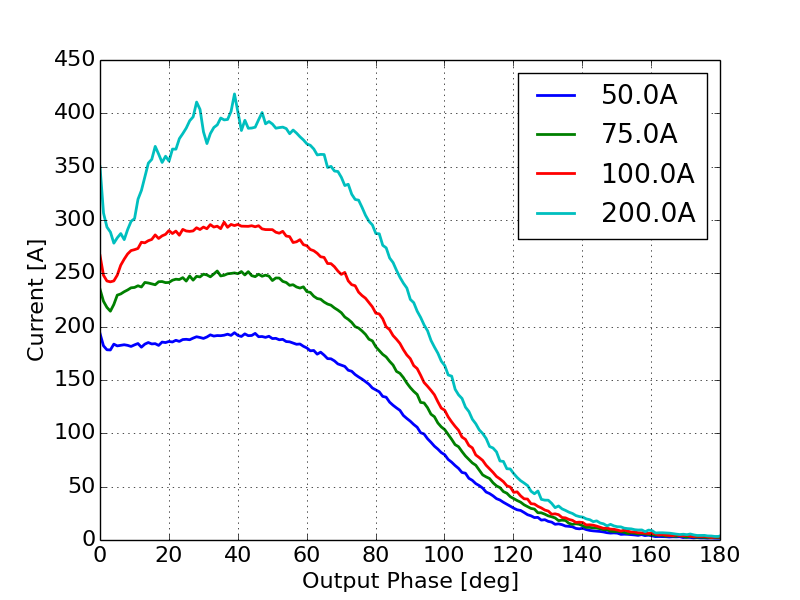}
        \caption{Beam current as a function of output phase for average injected currents ranging from 50 A to 200 A.}
    \end{subfigure}
    ~ 
    \begin{subfigure}[]{0.5\textwidth}
        \includegraphics[width=\textwidth]{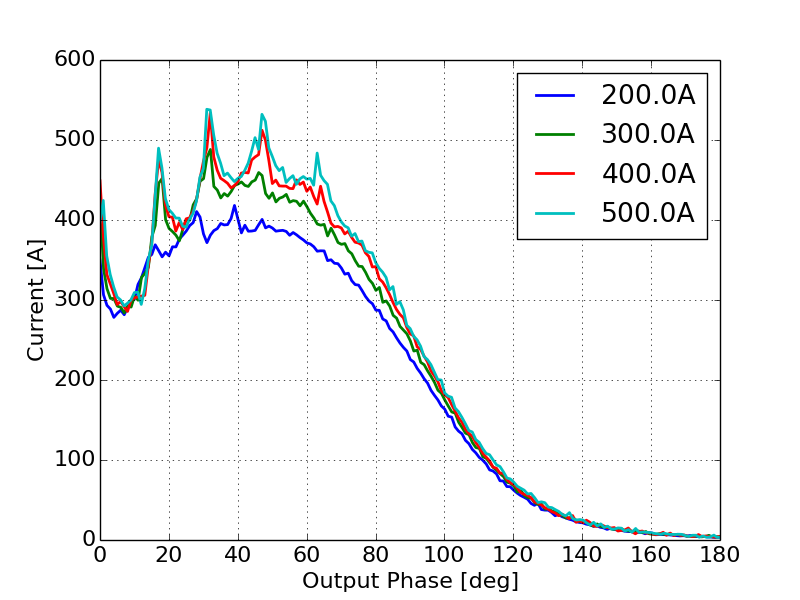}
        \caption{Beam current as a function of output phase for average injected currents ranging from 200 A to 500 A.}
    \end{subfigure}
    ~ 
    \begin{subfigure}[]{0.5\textwidth}
        \includegraphics[width=\textwidth]{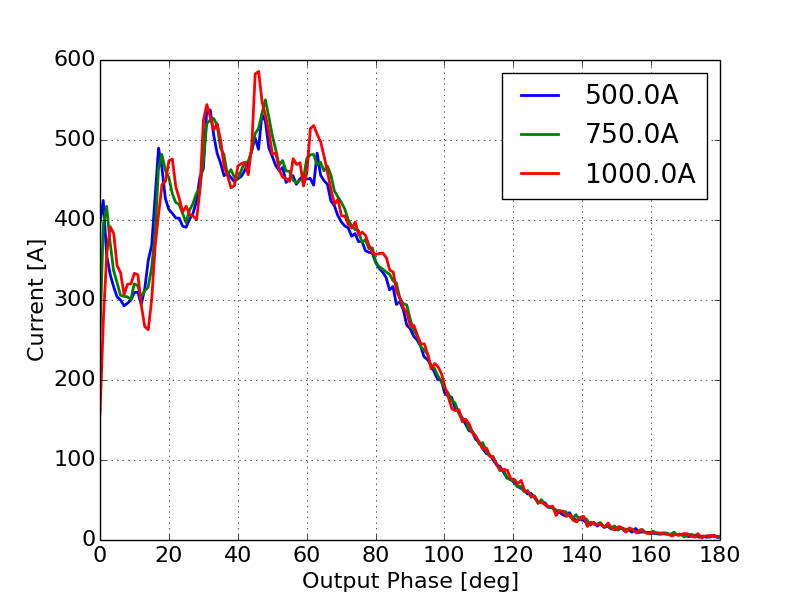}
        \caption{Beam current as a function of output phase for average injected currents ranging from 500 A to 1000 A.}
    \end{subfigure}
    \caption{Beam current as a function of output phase for average injected currents ranging from 50 A to 1000 A. $E_\mathrm{peak}$ was 40 MV/m and $\alpha$ is 3.33.}
\end{figure}

When the injected current is well above the space-charge limit this results in a virtual cathode and the formation of virtual cathode oscillations. These oscillations will modulate the field on the cathode which will in turn modulate the emission. The modulation of the emission will create an initial bunching of the beam. Additionally, the rate of acceleration off the cathode will be small due to the low fields on the cathode. This will result in a ballistic bunching that correlates with the field on the cathode as seen in Section III. Because the longitudinal space-charge field decays quickly into the gun, the net acceleration is only partially affected but there will be some energy modulation on the output beam correlated with the current modulation. To illustrate this point we computed the difference in final energy as a function of phase between simulations with high current and the 100mA simulation, Figure 16.

\begin{figure}[h]
\centering
\includegraphics[width=0.5\textwidth]{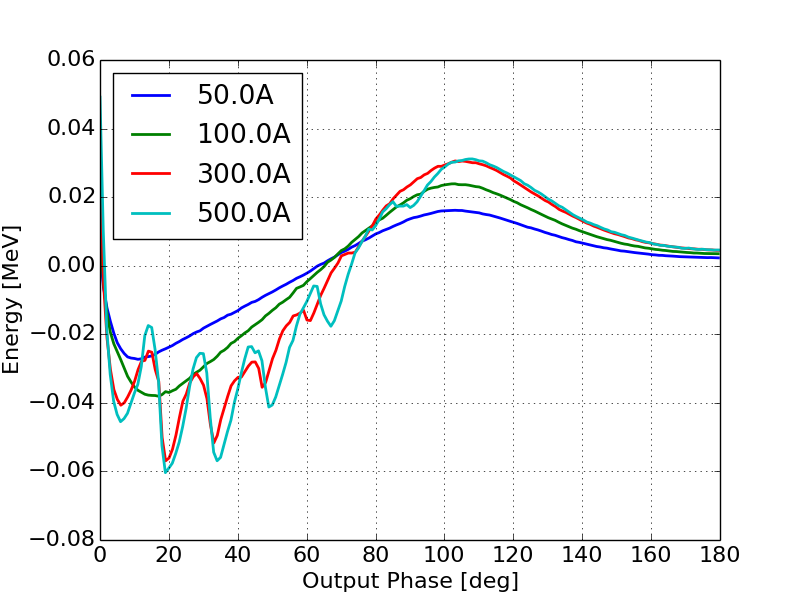}
\caption{Difference in final energy between simulations at high current and the 100mA simulation showing an energy modulation in the output beam caused by modulation of the cathode field. $E_\mathrm{peak}$ was 40 MV/m and $\alpha$ is 3.33.}
\end{figure}

Figure 17 shows the field on the cathode as a function of phase during the emission period of the gun as the current increases from 50 A to 1000 A. Figure 17a shows the initial onset of space-charge limited effects with the a very clear oscillation in the field during the first half of the emission period. As the space-charge intensity increases this oscillation becomes more prominent and begins to dominate, Figure 17b. Once the gun is fully space-charge limited we see an oscillatory field that has a frequency and amplitude which are dependent on the injected current, Figure 17c.

 \begin{figure}[h]
    \centering
    \begin{subfigure}[]{0.5\textwidth}
        \includegraphics[width=\textwidth]{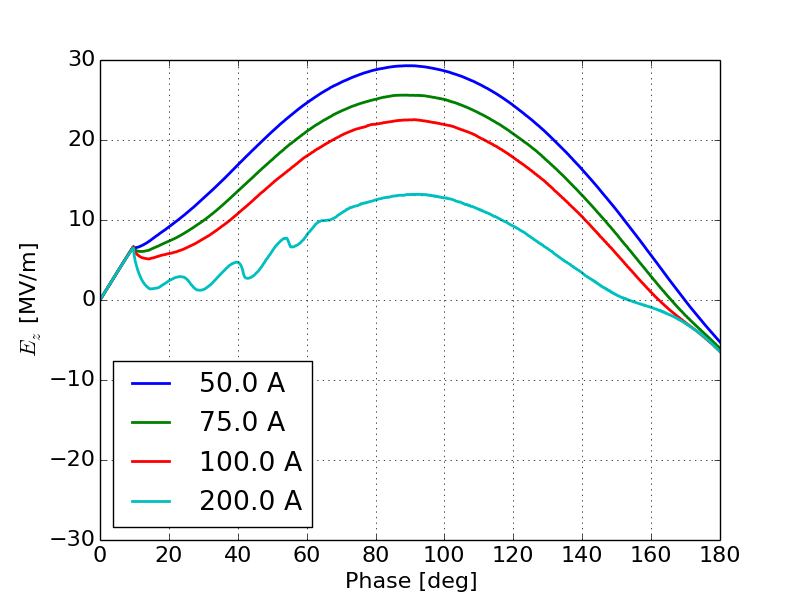}
        \caption{Net field on the cathode as a function of phase.}
    \end{subfigure}
    ~ 
    \begin{subfigure}[]{0.5\textwidth}
        \includegraphics[width=\textwidth]{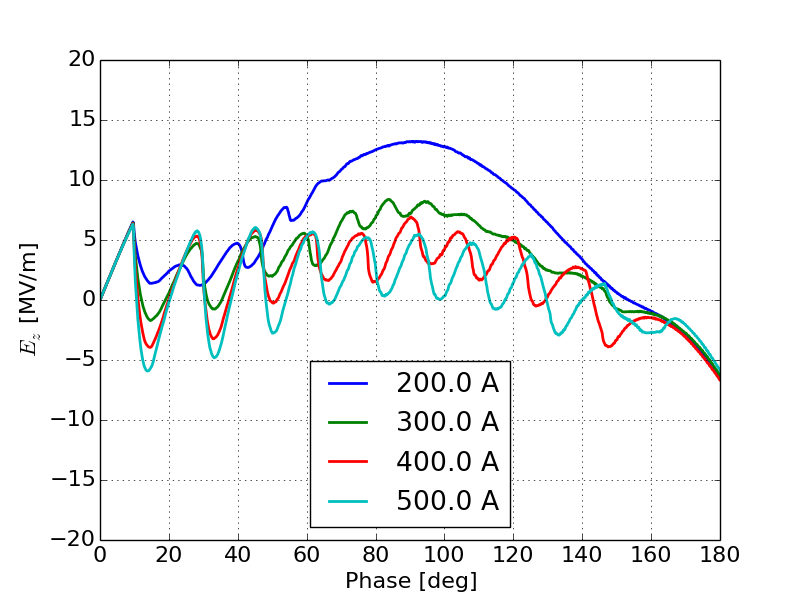}
        \caption{Net field on the cathode as a function of phase.}
    \end{subfigure}
    ~ 
    \begin{subfigure}[]{0.5\textwidth}
        \includegraphics[width=\textwidth]{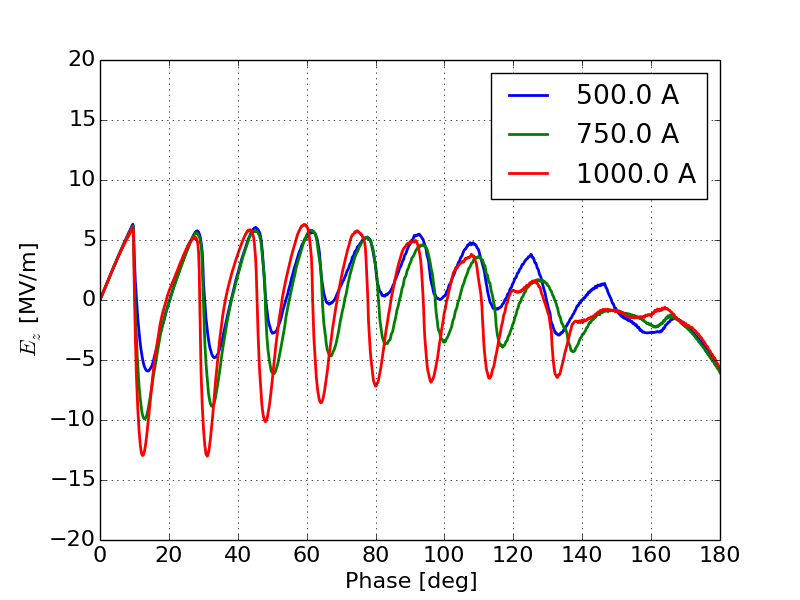}
        \caption{Net field on the cathode as a function of phase.}
    \end{subfigure}
    \caption{Net field on the cathode as a function of phase for average injected currents ranging from 50 A to 1000 A.  $E_\mathrm{peak}$ was 40 MV/m and $\alpha$ is 3.33.}
\end{figure}

This modulation of the field on the cathode will correspond directly to a modulation in the output current as the cathode will not emit when the field is less than zero. This can clearly be seen in the emission log from each of the simulations. Figure 18 shows histograms of the cathode emission beginning in the emission enhancement regime through the onset of space-charge effects and in the presence of fully space-charge limited emission.

\begin{figure}[h]
\centering
\includegraphics[width=1.0\textwidth]{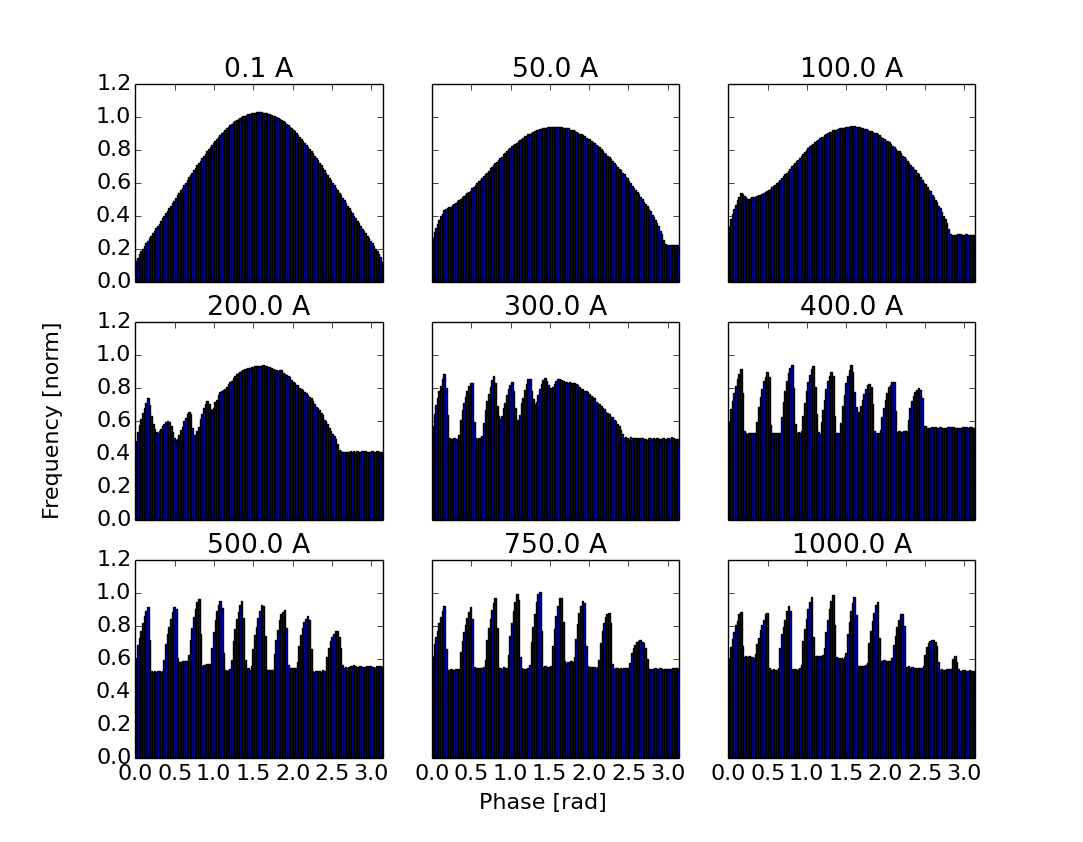}
\caption{Normalized histogram of the cathode emission log for average injected currents ranging from 0.1 A to 1000 A.  $E_\mathrm{peak}$ was 40 MV/m and $\alpha$ is 3.33.}
\end{figure} 

Here we can clearly see a modulation in the emission as the space-charge limit is approached. Note that there is a DC term in the emission log. This is due to the fact that thermionic cathodes will emit continuously even if beam is not accelerated off the cathode. Therefore when the field is negative we see emission corresponding to the average current injected into the simulation. If this modulation in the field on the cathode is a result of virtual cathode oscillations we would expect to see oscillation in the potentials near the cathode surface. To verify this we examine the scalar potential at several positions near the cathode as a function of time during the simulation. SPIFFE does output the scalar potentials however due to the simulation being on a finite grid it is difficult to obtain a uniform charge distribution while keeping the simulation computationally tractable. As a result the potential fields contain a large amount of numerical noise. The space-charge field however had significantly less noise. In order to compute the scalar potentials we used a moving average on the longitudinal field due to space-charge, then integrated to find the scalar potential at each point. The total scalar potential was then computed by adding the RF potential and the space charge potential. Figure 19 shows the scalar potential at several fixed locations very near the cathode as a function of phase during the simulation. 

 \begin{figure}[h]
    \centering
    \begin{subfigure}[]{0.5\textwidth}
        \includegraphics[width=\textwidth]{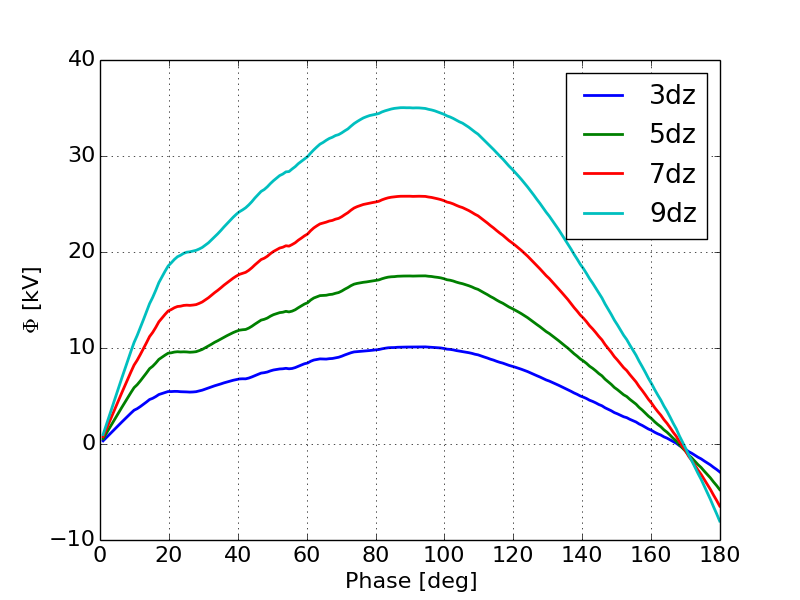}
        \caption{Net potential near the cathode surface as a function of phase for 200A of injected current.}
    \end{subfigure}
    ~ 
    \begin{subfigure}[]{0.5\textwidth}
        \includegraphics[width=\textwidth]{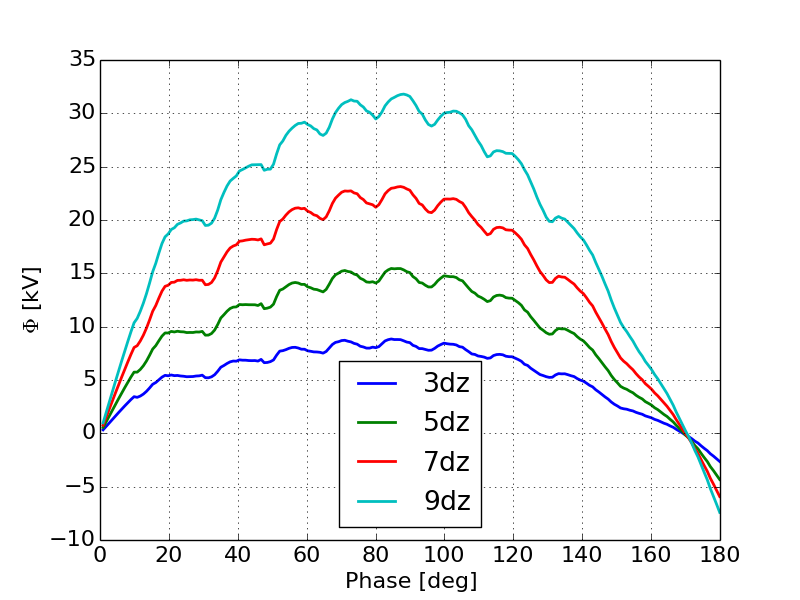}
        \caption{Net potential near the cathode surface as a function of phase for 500A of injected current.}    
        \end{subfigure}
    ~ 
    \begin{subfigure}[]{0.5\textwidth}
        \includegraphics[width=\textwidth]{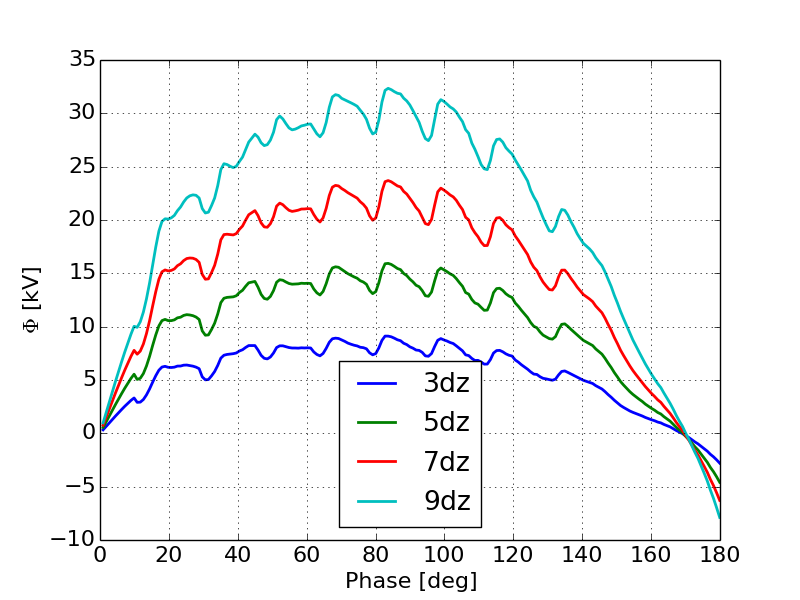}
        \caption{Net potential near the cathode surface as a function of phase for 1000A of injected current.}
    \end{subfigure}
    \caption{Electrostatic potential at several locations near the cathode as a function of phase for three values of injected current. Here dz represents the simulation mesh spacing in the longitudinal direction (approximately 500 $\mu m$).  $E_\mathrm{peak}$ was 40 MV/m and $\alpha$ is 3.33.}
\end{figure}

Here we see the clear formation of oscillations in the potential field near the cathode as the space-charge intensity increases. Figure 19a shows that for a relatively low current, the scalar potential does reach a local minimum with some small oscillations. As the space-charge intensity increases the oscillations become more prominent, Figure 19b. Finally when the gun is fully space-charge limited we see very clear  oscillations that mimimc the relaxation oscillations typically observed in DC guns [21]. Next we computed a histogram of the particle positions in the simulation weighted by their momentum at each time-step in the simulation for the 1 kA case. The observation of a local minimum in the histogram near the cathode indicates the formation of a virtual cathode. Additionally oscillations in the position of this minimum confirms that we are seeing a virtual cathode oscillation. Figure 20 shows the location of the local minimum in the histogram as a function of phase during the first half of the RF period. 

\begin{figure}[h]
\centering
\includegraphics[width=0.5\textwidth]{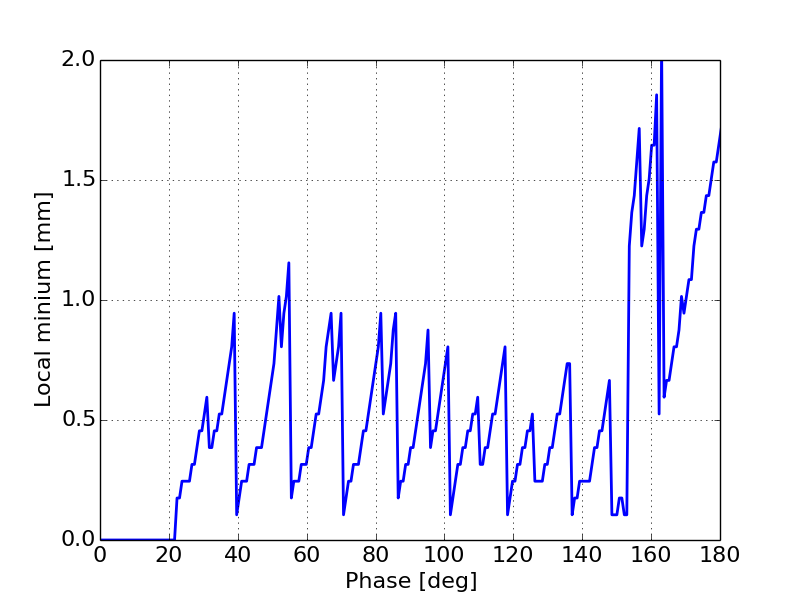}
\caption{Minimum in the weighted histogram as a function of phase during the first half of the simulation for an injected current of 1000 A.}
\end{figure} 

Here we see a very clear oscillation in the location of the local minimum during the simulation. Unfortunately, due to the large external fields the location of the potential depression is very close to the cathode. In order to resolve this in simulation to show conclusively that we are seeing virtual cathode oscillations very high resolution simulations are needed which require extensive computational resources. A more detailed investigation focusing on this behavior should be explored in the future. 

\section{Conclusions} 
This paper has increased the general understanding of how continuous emission in a thermionic RF gun will result in nonlinearities in the current distribution that cannot necessarily be compensated for down-stream. This represents a significant step forward towards understanding the fundamental limitations of these electron sources and how they can be applied to the next generation of accelerators. We have developed analytical models for the transmission efficiency and shown good comparison with simulation. Additionally we have characterized how different peak fields and gap lengths will affect the current distribution both through field enhanced emission and through bunch compression. Finally we have shown in detail how the onset of space-charge limited effects will impact the output current of the gun and how the onset of space-charge limited emission brings with it the formation of virtual cathode oscillations and as a result a modulation in the output current. Follow on studies will investigate how these current distributions evolve in multi-cell guns as well as how this longitudinal beam profile will impact the emittance produced by these guns.

\end{document}